\documentclass[prl,twocolumn,superscriptaddress,showpacs,amsmath,amssymb]{revtex4-1}

\usepackage{graphicx}
\usepackage{indentfirst}
\usepackage{psfrag}
\usepackage{epsfig}
\usepackage{amsmath}
\usepackage{amssymb}
\usepackage{bm}

\def\be{\begin{eqnarray}}
\def\ee{\end{eqnarray}}

\def\E{{\bf E}}

\def\p{{\bf p}}
\def\m{{\bf m}}

\begin{document}
  \title{Negative scattering asymmetry parameter for dipolar particles: \\ 
  Unusual reduction of the transport mean free path and radiation pressure
   }
\author{R. G\'omez-Medina} 
\affiliation{Departamento de F\'{\i}sica
de la Materia Condensada and Instituto ``Nicol\'as Cabrera'', Universidad Aut\'{o}noma de Madrid, 28049
Madrid, Spain.}
\author{L. Froufe-P\'erez} 
\affiliation{Departamento de F\'{\i}sica
de la Materia Condensada and Instituto ``Nicol\'as Cabrera'', Universidad Aut\'{o}noma de Madrid, 28049
Madrid, Spain.}
\author{M. Y\'epez}
\affiliation{Departamento de F\'{\i}sica
de la Materia Condensada and Instituto ``Nicol\'as Cabrera'', Universidad Aut\'{o}noma de Madrid, 28049
Madrid, Spain.}
\author{F. Scheffold}
\affiliation{Department of Physics, University of Fribourg,  Chemin
du Muse 3, 1700 Fribourg, Switzerland.}
 \author{M. Nieto-Vesperinas}
 \email{mnieto@icmm.csic.es} \affiliation{Instituto de Ciencia de Materiales de Madrid, Consejo
Superior de Investigaciones Cientificas, Campus de Cantoblanco,
Madrid 28049, Spain.}
\author{J. J. S\'aenz}
\email{juanjo.saenz@uam.es} \affiliation{Departamento de F\'{\i}sica
de la Materia Condensada and Instituto ``Nicol\'as Cabrera'', Universidad Aut\'{o}noma de Madrid, 28049
Madrid, Spain.}

\begin{abstract}
We establish a relationship between the electric magnetic dipole interaction force from a plane wave on a small magnetodielectric particle, the transport cross-section and the scattering asymmetry parameter, $g$. In this way, we predict negative $g$  that  minimize the transport mean free-path below  values of the scattering mean free path of a dilute suspension of both perfectly reflecting spheres as well as of those that satisfy the so-called Kerker conditions, like high permittivity dielectric ones.
\end{abstract}  \vspace{1cm}
\date{\today}
\pacs{05.60.Cd, 42.25.Dd, 42.55.-f, 42.25.Bs}
\maketitle









Propagation of light and image formation in turbid media has long been a subject of great interest \cite{Intro1} and constitutes the core of powerful techniques with countless applications including biomedical imaging \cite{Intro2a} and dynamic spectroscopy  techniques \cite{Intro2}, characterization of composite materials and complex fluids \cite{Intro3}, remote sensing or telecommunications \cite{Intro4}  to mention a few. 
Our current understanding of the diffusive transport through non-absorbing media is based on the knowledge of two key quantities: the transport and scattering mean free paths (mfp). The scatter density and cross section define the scattering mfp, $\ell_s$. The relevant scattering length for diffusive light power transport is the transport mfp, $\ell^*$. Both quantities are connected by the scattering asymmetry parameter $g$ defined \cite{Vandehulst, Bohren}  as the average of the cosine of the scattering angle, $g \equiv \langle \cos \theta \rangle$ with 
$\ell^* = \frac{\ell_s}{  1-g} . \label{mfp1}$.    
$\ell^*$ is usually equal to or larger than $\ell_s$ i.e. $g$ is positive. For instance, the isotropic Rayleigh scattering of small particles lead to $g \sim 0$ while  Mie particles (or human tissue) \cite{Bohren}  scatter strongly in the forward direction (small scattering angles) and hence $g \sim 1$.  However, very recently it has been shown that subwavelength spheres made of non-absorbing dielectric material with relatively large refractive index produce anisotropic angular distributions of scattered intensity \cite{Etxarri, Nieto_Opex,Nieto_JOSA,Nanophot}. As we will show here, these particles can present negative $g$ values in specific wavelength regions, i.e. a random dispersion of such particles 
will show the unusual characteristic of having $\ell^* < \ell_s$,
even in the absence of positional correlations.

The transport mean free path  can be strongly modified by the presence of short range structural order in the system. \cite{Maret,Cefe}
Positional correlations  usually lead to positive $g$ values, i.e.  to $\ell^*$ values significantly larger than $\ell_s$  which are responsible, for example, for the relatively large conductivity of disordered liquid metals  \cite{metal} or the transparency of the cornea to visible light \cite{cornea}. However, short range order can also lead to
negative values of the asymmetry parameter  as it has been recently shown in experiments in colloidal liquids \cite{Frank1} and amorphous photonic materials \cite{Frank2}. 
These negative values, observed at specific wavelength regions, has been associated \cite{Frank1,Frank2} to enhanced  backscattering at Bragg-like matching resonances \cite{birds}. 

The unusual observation of negative $g$ factors has been limited  to systems with 
appropriate short range correlation between scatters. While it is frequently argued that the scattering from Mie spherical particles lead to $g>1$, in this Letter we show that non-absorbing Mie spheres of relatively low refraction index present negative $g$ factors in specific spectral ranges. Interestingly, we will show  that small particles, whose scattering may be completely described by a dipolar response both to the electric and magnetic fields, may also present negative asymmetry parameters. 
As we will see,  there is a close relationship between transport parameters of a dilute suspension of dipolar particles and the theory of optical forces on magnetodielectric small particles 
 \cite{Chaumet_magnet,Nieto_Opex,Nieto_JOSA}, in which it has been shown that, in addition
   to the force due to the electric and magnetic induced dipoles, there is an additional component due to the interaction between both of them which was associated to the angular distribution of  scattered intensity \cite{Nieto_Opex,Nieto_JOSA,Nanophot}.  Maxima and (negative) minima of the $g$ factor are obtained at the  so-called Kerker's conditions \cite{Nieto_JOSA,Kerker} of  zero-backward or almost zero-forward differential scattering cross sections (DSCS).
    These conditions can be satisfied by small dielectric  particles of high refractive index (e.g. of Si or Ge), that have recently been shown to behave as magnetodielectric,  \cite{Etxarri,Nanophot} 
      i.e. whose scattering is effectively dipolar, being well characterized by the Mie coefficients $a_1$ and $b_1$, both being of comparable  strength. A dilute suspension of such particles  near the almost zero-forward scattering condition, will minimize the transport mean free-path below the scattering mean free path.

\begin{figure}
\begin{center}
\includegraphics[width=8cm]{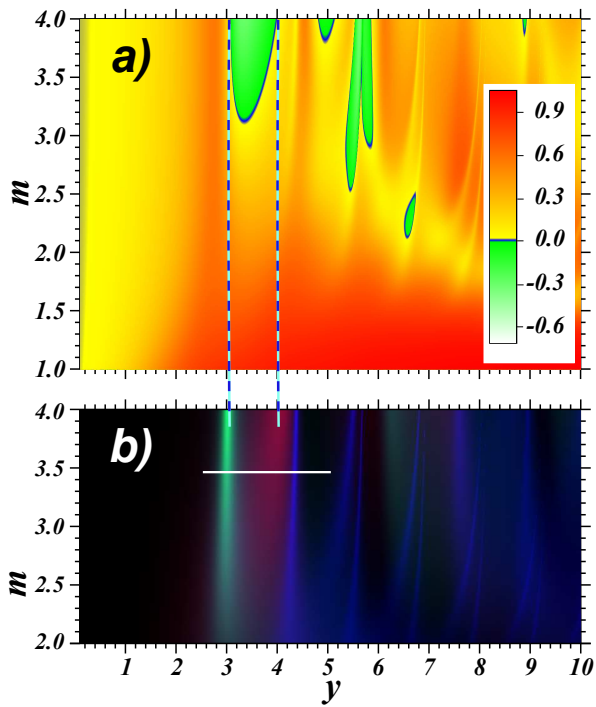}
\caption{a) Color map of the $g$ factor for spherical absorptionless
particles as a function of their refractive index $m$ and size parameter $y=mka$. As seen in
the attached scale, green areas correspond to negative values of
$g$. b) Color map of the sphere scattering cross section. Red corresponds
to dominant electric dipole contributions to the scattering cross
section. Green corresponds to dominant magnetic dipole
contributions, while blue sums up all higher order multipole terms.
 Vertical dashed lines coincide with $y$-parameter for
maximum electric dipole contribution (right vertical line) and 
maximum magnetic dipole contribution (left vertical line). 
The white horizontal line  at $m\approx 3.5$, (which corresponds  to a Silicon sphere), represents the y-range covered by figure 2.}
\label{force}
\end{center}
\end{figure}
The {\it asymmetry factor} $g=\langle \cos \theta \rangle$ is defined  \cite{Vandehulst, Bohren} as the average of the cosine of the scattering angle, $\theta$, over the particle differential scattering cross section distribution
$d\sigma_s/d\Omega$:
\begin{equation}
g=\langle \cos \theta \rangle=\frac{\int\frac{d\sigma_s}{d\Omega}\cos\theta
d\Omega}{\int\frac{d\sigma_s}{d\Omega} d\Omega}=\frac{\int\frac{d\sigma_s}{d\Omega}\cos\theta
d\Omega}{\sigma_s}.
\label{tetamean}
\end{equation}
where  $\sigma_s$ the scattering cross section.
Let us compute the $g$ factor for  a dielectric dipolar sphere
of radius $a$ and real refractive index $m_p$ immersed in an
arbitrary lossless medium with  relative dielectric permittivity
$\epsilon$ and magnetic permeability $\mu$.
 For spherical particles, the $g$ factor does not depend on the polarization of the incident light and can be expressed in terms of the ``Mie'' coefficients  $a_n$ and $b_n$ 
(see Sec. 4.5 in Ref.  \cite{Bohren}). In Fig. \ref{force}a we show the $g$ factor map for a non-absorbing Mie sphere  as a function of the relative refractive index $m=m_p/\sqrt{\epsilon \mu}$ and the size parameter $y \equiv  m (2\pi a /\lambda)$ calculated from the full Mie expansion. Since usually non-absorbing materials present low refractive index ($m \lesssim 1.5$) in the infrared (IR) and visible frequency ranges, negative $g$ factors in Mie particles were not expected. However, as it can be seen in Fig. \ref{force}a, the $g$ map shows regions of negative $g$ for relatively low refraction index ($m \gtrsim 2$) relevant for semiconductor particles made of  Silicon ($m \approx 3.5$) or Germanium ($m\approx4$)  in the infrared and telecom wavelengths. This is one of the main results of the present work. The corresponding scattering cross section map for the same spheres (as calculated in Ref. \cite{Etxarri}) is plotted in Fig. \ref{force}b.  For $m$ values larger than $\approx 2$, the region between the magnetic (green colors in Fig. \ref{force}b) and electric (red colors in Fig. \ref{force}b) dipolar resonances  presents a well defined region of negative asymmetry parameter while the scattering is perfectly described by the first two dipolar terms in the Mie expansion \cite{Etxarri}.

The asymmetry factor $g$ and the  seemingly unrelated problems of transport mean free path and optical forces are now tied together by the close relation between power and momentum transfer, i.e.  by the definition of transport, $\sigma^*$, and radiation pressure, $\sigma^{(pr)}$, cross sections.
 \cite{Vandehulst, Bohren, Irvine} 
The transport cross section $\sigma^{\ast}$ of a particle is expressed in terms of $\sigma_s$ as \cite{Intro1,Intro2}:
\begin{equation}
\sigma^{\ast}=\int\frac{d\sigma_s}{d\Omega}(1-\cos\theta) d\Omega= \sigma_s (1-g), \label{sigmatrans}
\end{equation}
For a dilute suspension of optically uncorrelated particles with density $\rho$, the transport mfp, $l^{\ast}=1/(\rho\sigma^{\ast})$, is related to the scattering mfp, $l_s=1/(\rho\sigma_s)$,
through the aforementioned relationship: $\ell^* = \frac{\ell_s}{  1-g} $. On the other hand, the radiation pressure cross section is
customarily defined as \cite{Vandehulst} $\sigma^{(pr)} =\sigma^{(ext)}- <\cos\theta>\sigma_s=\sigma_a+\sigma^{\ast}, $
where $\sigma^{(ext)}=\sigma_s+\sigma_a$, $\sigma_a$
being the absorption cross section. In absence of absorption, there is no difference between transport and radiation pressure cross sections,
\begin{equation}
\sigma^{\ast}=\sigma^{(pr)}= \sigma_s (1-g). \label{fp}
\end{equation}
Hence there is  a direct relation between transport 
 quantities and the forces from an incident plane wave on a dielectric sphere.

In order to get a deeper physical insight on the influence of the electric-magnetic dipole force in $\ell^*$, it is interesting to derive the explicit expressions connecting both transport and radiation pressure with the $g$ factor for the simplest and most important case of dipolar particles.  
Let us consider  such a  particle whose dipolar electric, $\p$, and magnetic, $\m$, moments are related to the external polarizing fields through  $\p = \epsilon_0 \epsilon \alpha_e \E$ and $\m = (\alpha_m/\mu_0\mu) \bm{B}$. The dynamic polarizabilities,  $\alpha_e$ and $\alpha_m$, 
 that characterize the dipole excitation can be  expressed in terms of  the Mie
coefficients $a_1$ and $b_1$ as  \cite{Vandehulst, Bohren}: 
$
 \alpha_e =  i  a_1 (6\pi/ k ^3) $ and
$\alpha_m = i  b_1 (6\pi/ k ^3), $
 ($k$ is the wavenumber: $k =
\sqrt{\epsilon \mu} \ \omega/c$).

The
differential scattering cross section, averaged over the incident
polarizations, is \cite{Nieto_Opex,Nieto_JOSA,Nanophot}:
\begin{align}
\frac{d\sigma_s(\theta)}{d\Omega}  = \frac{k ^4}{16 \pi}   \Big(&
\left(\left| \alpha_e\right|^2+\left|\alpha_m\right|^2
\right) (1+ \cos^2 \theta)  \nonumber \\ &+
 2   \text{Re}(\alpha_e \alpha_m^* ) \cos \theta \Big).
\label{difcross2}
\end{align}
From Eqs. (\ref{difcross2}) and (\ref{tetamean})
 \be
 g=\frac{\text{Re}[
\alpha_e \alpha_{\text{m}}^{\ast}]}{ |  \alpha_e|^2+|
 \alpha_m|^2 } \ , \label{factorg}
  \ee
 which shows that, for a dipolar particle,  $|g|\leq 1/2$. 
 Notice also that from Eqs.
(\ref{difcross2})  and (\ref{factorg}) this asymmetry factor may be
expressed as
 \be
 g=\frac{1}{2}\left[\frac{
 \frac{d\sigma^{(s)}}{d\Omega}(0^\circ)-
 \frac{d\sigma^{(s)}}{d\Omega}(180^\circ)}
 {\frac{d\sigma^{(s)}}{d\Omega}(0^\circ)+
 \frac{d\sigma^{(s)}}{d\Omega}(180^\circ)}\right].
 \label{factorgdsigma}
  \ee
On the other hand, in absence of absorption, the time averaged force exerted on the dipolar  particle
by a time harmonic incident plane wave, $\E = 
\E_0 e^{i \bf{k} \cdot {\bf r}}$  is all radiation pressure
\cite{Nieto_Opex,Nieto_JOSA} and reads:
\begin{align}
\langle {\bf F}\rangle &= <{\bf F}_{e}> + <{\bf F}_{m}>+ <{\bf F}_{e-m}>
\label{threeterms}
\\ &= \frac{\epsilon_0 \epsilon}{2} |\E_0|^2  \left\{k \  \text{Im} \left(\alpha_e+ \alpha_m\right)-\frac{k^4}{6\pi}\text{Re}[
\alpha_{e} \alpha_m^{\ast}]\right\}
 \nonumber \\ &= \frac{\epsilon_0 \epsilon}{2} |\E_0|^2 \sigma^{(pr)} \frac{{\bf k}}{k} 
\end{align}
The last term  in Eq. (\ref{threeterms}), $<{\bf F}_{e-m}>$, due to the interaction between electric and magnetic dipoles \cite{Chaumet_magnet,Nieto_Opex,Nieto_JOSA},
 was in \cite{Nieto_JOSA} associated to the asymmetry in the scattered intensity distribution, [cf. the last term in Eq.  (\ref{difcross2})]
even though it was not explicitly related to $g$. We next show that they are proportional.  Notice that the moduli of the first
two terms 
 $<{\bf F}_{e}>$ and $<{\bf F}_{m}>$, corresponding to the forces on the induced pure electric and magnetic dipoles can be written as 
 \begin{eqnarray}
 <F_{e}> + <F_{m}> = \frac{\epsilon_0 \epsilon}{2} |\E_0|^2 \sigma^{(ext)} = \frac{\epsilon_0 \epsilon}{2} |\E_0|^2 \sigma_s
\end{eqnarray}
where the last  equality holds for non-absorbing particles. 
while the interference term
 \begin{eqnarray}
 <F_{e-m}>  &=& - \frac{\epsilon_0 \epsilon}{2} |\E_0|^2 \sigma_s g
\end{eqnarray}
We then have a formal result for the total force
\be
\langle F \rangle = ( <F_{e}> + <F_{m}> ) (1-g)
\ee
which is the force analogue of Eq. (\ref{fp}). We can summarize the above discussion in a single expression:
\be
1-g = \frac{\sigma^*}{\sigma_s} = \frac{\langle F \rangle}{  <F_{e}> + <F_{m}> } = \frac{\ell_s}{\ell^*} \label{summ}
\ee
Equation (\ref{summ}) is another main results of this work.
When the particle is non absorbing,  $1-g$
becomes just the ratio between the magnitudes of the total force and the sum of the pure electric dipole forces.
This quantifies in a specific way the nature of the interaction
force component in terms of the asymmetry forward-backward of the
angular distribution of scattered intensity by the particle. It also
establishes the connection between these forces and the transport and scattering mfp's.

If no restrictions are imposed on $\alpha_e$ and
$\alpha_m$, and hence one may consider them in Eq. (\ref{factorg}) as
independent variables, it is straightforward to see from this
equation that $g$ takes on extreme values when either $
\alpha_{e}=\alpha_{m}$, ($g$ then being a maximum: $g=1/2$), or
when $\alpha_{e}=- \alpha_{m}$, ($g$ then being a
minimum: $g=-1/2$). The first condition corresponds to the so-called 
{\it first Kerker condition} and has been discussed in the context of scattering from
a special case of magnetodielectric particles  \cite{Kerker,Nieto_JOSA}. 
 These particles lead to {\it zero backward differential scattering cross section}
 and have $g=1/2$,
i.e. $l^{\ast}=2l_s$ (notice that forward scattering would
correspond to $g=1$ and $l^{(\ast)}=\infty$). The second condition 
($\alpha_{e}=- \alpha_{m}$) minimize the  {\it scattered intensity in the forward direction}
can only be fulfilled aproximatelly since the imaginary part of the polarizabilities must be always positive 
as required from causality \cite{Nieto_JOSA}. In the quasistatic approximation, they produce {\it zero forward
scattered power} \cite{Kerker}) with an asymmetry factor $g \approx -1/2$,
which means that $l^{\ast}\approx (2/3) l_s$ (notice that strong
backscattering would correspond to $g=-1$ and
$l^{\ast}=(1/2)l_s$). 
 Thus, for a diluted
suspension of arbitrary non-absorbing dipolar particles, Eq. (\ref{factorg}) and the Kerker conditions impose
the following limits to the mfp:
\be
0.66 \ \ell_s \lesssim  \ell^* \le 2 \ \ell_s
\ee
\begin{figure}
\begin{center}
\includegraphics[width=8cm]{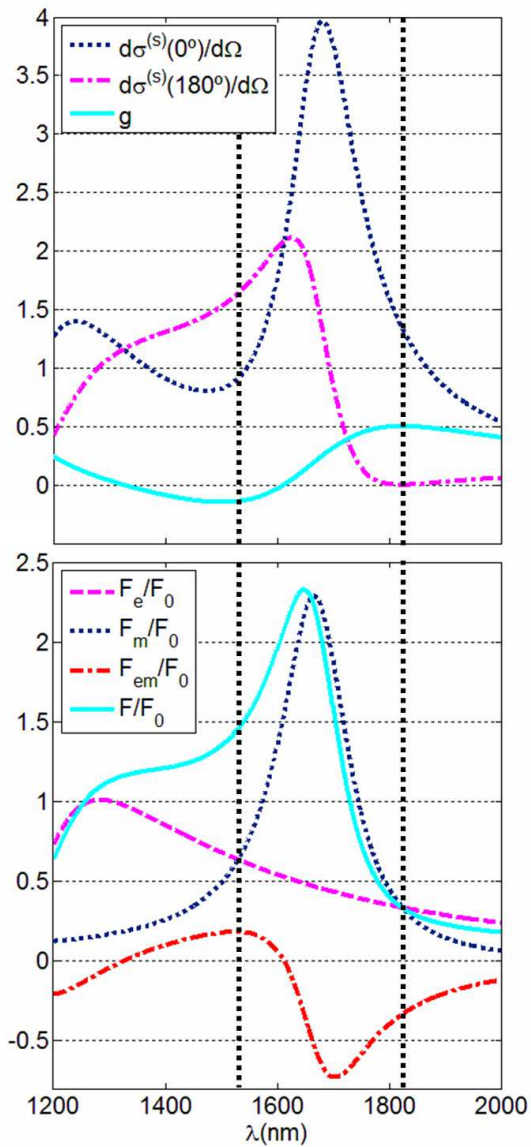}
\caption{(a)Forward and backward differential scattering cross
section, and asymmetry factor versus the wavelength, for a Silicon
spherical particle of radius $a = 230nm$, $\epsilon = 12$. (b)
Different contributions to the total radiation pressure, versus the
wavelength, for the same particle. Normalization is done by
$F_0=4 \pi a^3 k |\E_0|^2/2$. The vertical lines mark, from right
to left, the first and second Kerker conditions.} \label{Fgfactor}
\end{center}
\end{figure}
While the results above are of general application, it is interesting to discuss two specific examples that 
could be realized experimentally.
As a  direct consequence of the discussion  above, 
 perfectly  conducting dipolar spheres \cite{Jackson},
for which $\alpha_{e}^{(0)}=-2 \alpha_{m}^{(0)}$,
have an asymmetry factor which is negative. In particular in
the quasistatic limit this factor  becomes $g=-0.4$, which implies
$\ell^* \approx 0.7 \ell_s$ for a random dispersion of small, perfect  conducting spheres.
This result may be specially relevant in the THz regime where metals can be often 
considered as perfect conductors \cite{THz}.

Si spheres with radius
$a=230 nm$ have been proven to behave as dipolar magnetodielectric
particles with a strong magnetic dipole response in the near
infrared. In Fig.\ref{Fgfactor} we show the forward and backward
differential scattering cross section and the asymmetry factor $g$
as well as the variation of  $<{\bf F}_{e-m}>$, $<{\bf F}_{e}>$ and
$<{\bf F}_{m}>$  for a Si sphere of radius $230 nm$. Notice when the
first Kerker condition is fulfilled, $g=1/2$ and
$\frac{d\sigma^{(s)}}{d\Omega}(180^\circ)=0$ (Fig.\ref{Fgfactor}.a)
and $F_e=F_m=-F_{em}=F$ (Fig.\ref{Fgfactor}.b).

Nevertheless, if one imposes restrictions on $\alpha_e$ and
$\alpha_m$, then other situations appear. From
Eq.(\ref{factorgdsigma}) one sees at once that $g$ is maximum and
equal to $1/2$ where $\frac{d\sigma_s}{d\Omega}(180^\circ)$ is
zero and it has a minimum value at that wavelength where
$\frac{d\sigma_s}{d\Omega}(0^\circ)$ is minimum. This minimum
value of $g$ being also negative if
$\frac{d\sigma_s}{d\Omega} (0^\circ)<
\frac{d\sigma_s}{d\Omega}(180^\circ)$. This is illustrated in
the above mentioned Si sphere of radius $230 nm$. As seen in
Fig.\ref{Fgfactor}.a , at $\lambda=1530nm$  $g$ has a minimum equal
to {\bf -0.15} which corresponds to the minimum forward DSCS,
whereas where the first Kerker condition holds, then being zero the
backscattering cross section, the $g$ factor has a maximum equal to
$1/2$.

In conclusion, we have demonstrated that, surprisingly and without further assumptions on collective interactions,  dilute suspensions of dipolar semiconductor spheres,   like e.g Si and Ge, have an optical frequency range in which their scattering asymmetry  parameter is negative, hence they acquire a transport mfp smaller than their scattering mfp. This is made possible by the magnetodielectric nature of these particles and the consequent electric-magnetic dipole interference which, in addition, leads to a simple relation between the electric-magnetic interaction photonic force and the  asymmetry factor. This also applies to perfectly conducting spheres at longer wavelengths.


This work was supported by the Spanish MEC through the Consolider
\textit{NanoLight} (CSD2007-00046)
 and FIS2009-13430-C01-C02 research grants and by the Comunidad de Madrid \textit{Microseres-CM} Project (S2009/TIC- 1476). M.Y.  thanks  CONACYT for a  postdoctoral grant (Ref: 000000000162768).

\end{document}